\documentclass[a4paper,twoside]{article} 
  \usepackage{array}                     %
  \usepackage{amsmath}                   %
  \usepackage{graphicx}                  
  \usepackage{amssymb}                   
  \usepackage{hyperref}                  %
  \usepackage{amsthm}                    %

  \theoremstyle{plain} \newtheorem{Thm}{Theorem}
  \theoremstyle{definition} \newtheorem{Def}{Definition}
  \theoremstyle{remark} \newtheorem{Rem}{Remark}

\begin{document}


\title{CACHING IN MULTIDIMENSIONAL DATABASES\footnote{Web: \href{http://www.pp.bme.hu/ee/2007_3/pdf/ee2007_3_06.pdf}{http://www.pp.bme.hu/ee/2007\_3/pdf/ee2007\_3\_06.pdf}; Journal reference: Periodica Polytechnica Electrical Engineering, Vol. 51, Number 3-4, pp. 119-132, 2007; DOI: 10.3311/pp.ee.2007-3-4.06.}}

\author{Istv\'an {\sc Sz\'epk\'uti}\\
\\
ING Insurance Hungary Pte.\ Co.\ Ltd.\\
H-1068 Budapest, D\'ozsa Gy\"orgy \'ut 84/b, Hungary\\
e-mail: szepkuti@inf.u-szeged.hu}
\date{Received: September 12, 2006} \maketitle

\markboth{{\tiny I. SZ\'EPK\'UTI}}{{\tiny CACHING IN
MULTIDIMENSIONAL DATABASES}} \pagestyle{myheadings}

\begin{abstract}
One utilisation of multidimensional databases is the field of
On-line Analytical Processing (OLAP). The applications in this
area are designed to make the analysis of shared multidimensional
information fast \cite{PendseB}.

On one hand, speed can be achieved by specially devised data
structures and algorithms. On the other hand, the analytical
process is cyclic. In other words, the user of the OLAP
application runs his or her queries one after the other. The
output of the last query may be there (at least partly) in one of
the previous results. Therefore caching also plays an important
role in the operation of these systems.

However, caching itself may not be enough to ensure acceptable
performance. Size does matter: The more memory is available, the
more we gain by loading and keeping information in there.

Oftentimes, the cache size is fixed. This limits the performance
of the multidimensional database, as well, unless we compress the
data in order to move a greater proportion of them into the
memory. Caching combined with proper compression methods promise
further performance improvements.

In this paper, we investigate how caching influences the speed of
OLAP systems. Different physical representations (multidimensional
and table) are evaluated. For the thorough comparison, models are
proposed. We draw conclusions based on these models, and the
conclusions are verified with empirical data. In particular, using
benchmark databases, we show examples when one physical
representation is more beneficial than the alternative
one and vice versa.\\
\\
\emph{Keywords:} compression, caching, multidimensional database,
On-line Analytical Processing, OLAP.
\end{abstract}


\section{Introduction}

\subsection{Motivation}

Why is it important to investigate the caching effects in
multidimensional data\-bases?

A number of papers compare the different physical representations
of data\-bases in order to find the one resulting in higher
performance than others. For examples, see \cite{Graefe, Ray,
SzepkutiC, SzepkutiD, Tolani, ZhaoB}. However, many of these
papers either ignore the influence of caching or discusses this
issue very briefly.

As it will be shown later, the size of the buffer cache affects
the results significantly. Hence the thorough analysis of the
buffering is necessary in order to better understand what is the
real reason of the performance improvements.

\subsection{Results}

The results of this paper can be summarized as follows:
\begin{itemize}
    \item Two models are proposed to analyse the caching effects of
          the alternative physical representations of relations.
    \item With the help of the models, it is shown that the
          performance difference between the two representations
          can be several orders of magnitude depending on the size of
          the buffer cache.
    \item It is also demonstrated that the generally better
          multidimensional physical representation may become
          worse, if the memory available for caching is large enough.
    \item The models are verified by a number of experiments.
\end{itemize}

\subsection{Related Work}

In the literature, several papers deal with compressed databases:
For further details the reader may wish to consult \cite{Chen,
Kaser, O'Connell, Westmann, Wu}.

The paper of {\sc Westmann} \emph{et al.}\ \cite{Westmann} lists
several related works in this field. It also discusses how
compression can be integrated into a relational database system.
It does not concern itself with the multidimensional physical
representation, which is the main focus of our paper. They
demonstrate that compression indeed offers high performance gains.
It can, however, also increase the running time of certain update
operations. In this paper we will analyse the retrieval (or point
query) operation only, as a lot of On-line Analytical Processing
(OLAP) applications handle the data in a \emph{read only} or
\emph{read mostly} way. The database is updated outside working
hours in batch. Despite this difference, we also encountered
performance degradation due to compression when the entire
physical representation was cached into the memory. In this case,
at one of the benchmark databases (TPC-D), the multidimensional
representation became slower than the table representation because
of the CPU-intensive Huffman decoding.

In this paper, we use difference\,--\,Huffman coding to compress
the multidimensional physical representation of the relations.
This method is based on difference sequence compression, which was
published in \cite{SzepkutiD}.

{\sc Chen} \emph{et al.}\ \cite{Chen} propose a Hierarchical
Dictionary Encoding and discusses query optimization issues. Both
of these topics are beyond the scope of our paper.

In the article of {\sc O'Connell} \emph{et al.}\ \cite{O'Connell},
compressing of the data itself is analysed in a database built on
a triple store. We remove the empty cells from the
multidimensional array, but do not compress the data themselves.

When we analyse algorithms that operate on data on the secondary
storage, we usually investigate how many disk input/output (I/O)
operations are performed. This is because we follow the
\emph{dominance of the I/O cost} rule \cite{Garcia-Molina}. We
followed a similar approach in Section \ref{sec:IOCost} below.

The main focus of \cite{Chen2005} is the CPU cache. In our paper,
we deal with the buffer cache as opposed to the CPU cache.

{\sc Vitter} \emph{et al.}\ \cite{Vitter} describe an algorithm
for prefetching based on compression techniques. Our paper
supposes that the system does not read ahead.

{\sc Poess} \emph{et al.}\ \cite{Poess} show how compression works
in Oracle. They do not test the performance for different buffer
cache sizes, which is an important issue in this paper.

In \cite{Xi}, {\sc Xi} \emph{et al.}\ predict the buffer hit rate
using a Markov chain model for a given buffer pool size. In our
article, instead of the buffer hit rate, we estimate the expected
number of pages brought into the memory from the disk, because it
is proportional to the retrieval time. Another difference is that
we usually start with a cold (that is empty) cache and investigate
its increase together with the decrease in retrieval time. In
\cite{Xi}, the authors fix the size of the buffer pool and then
predict the buffer hit rate with the Markov chain model.

\subsection{Organisation}

The rest of the paper is organised as follows. Section
\ref{sec:PhyicalRepresentations} describes the different physical
representations of relations including two compression techniques
used for the multidimensional representation. Section
\ref{sec:IOCost} introduces a model based on the dominance of the
I/O cost rule for the analysis of the caching effects. An
alternative model is presented in Section \ref{sec:Alternative}.
The theoretical results are then tested in experiments outlined in
Section \ref{sec:Experiments}. Section \ref{sec:Conclusion} rounds
off the discussion with some conclusions and suggestions for
future study. Lastly, for the sake of completeness, a list of
references ends the paper.

\section{Physical Representations of Relations}
\label{sec:PhyicalRepresentations}

Throughout this paper we use the expressions `multidimensional
representation' and `table representation,' which are defined as
follows.

\begin{Def} Suppose we wish to represent relation $R$ physically. The
multidimensional (physical) representation of $R$ is as follows:

\begin{itemize}
    \item A compressed array, which only stores the nonempty cells, one
          nonempty cell corresponding to one element of $R$;
    \item The header, which is needed for the logical-to-physical position
          transformation;
    \item One array per dimension in order to store the dimension values.
\end{itemize}

The table (physical) representation consists of the following:

\begin{itemize}
    \item A table, which stores every element of relation $R$;
    \item A B-tree index to speed up the access to given rows of the
          table when the entire primary key is given. \hfill $\square$
\end{itemize}
\end{Def}

In the experiments, to compress the multidimensional
representation, difference\,--\,Huffman coding (DHC) was used,
which is closely related to difference sequence compression (DSC).
These two methods are explained in the remainder of this section.

\emph{Difference sequence compression.} By transforming the
multidimensional array into a one-dimensional array, we obtain a
sequence of empty and nonempty cells:
\begin{equation*}
(E^*F^*)^*
\end{equation*}
In the above regular expression, $E$ is an empty cell and $F$ is a
nonempty one. The difference sequence compression  stores only the
nonempty cells and their logical positions. (The logical position
is the position of the cell in the multidimensional array before
compression. The physical position is the position of the cell in
the compressed array.) We denote the sequence of logical positions
by $L_j$. This sequence is strictly increasing:
\[
L_0 < L_1 < \dots < L_{N-1}.
\]
In addition, the difference sequence $\Delta L_j$ contains smaller
values than the original $L_j$ sequence. (See also Definition
\ref{def:Lj} below.)

The \emph{search algorithm} describes how we can find an element
(cell) in the compressed array. During the design of the data
structures of DSC and the search algorithm, the following
principles were used:

\begin{itemize}
    \item We compress the header in such a way that enables quick decompression.
    \item It is not necessary to decompress the entire header.
    \item Searching can be done during decompression, and the decompression
          stops immediately when the header element is found or when it is
          demonstrated that the header element cannot be found (that is, when
          the corresponding cell is empty).
\end{itemize}

\begin{Def} \label{def:Lj} Let us introduce the following notations.\\
$N$ is the number of elements in the sequence of logical positions ($N > 0$);\\
$L_j$ is the sequence of logical positions ($0 \leqq j \leqq N - 1$);\\
$\Delta L_0 = L_0$;\\
$\Delta L_j = L_j - L_{j-1}\ (j = 1, 2, \dots, N - 1)$;\\
The $D_i$ sequence ($D_i \in \lbrace 0, 1, \dots , \overline{D}
\rbrace,\ i = 0, 1, \dots, N - 1$) is defined as follows:
\begin{equation*}
D_i = \left \lbrace
\begin{array}{ll}
\Delta L_i, & $if $\Delta L_i \leqq \overline{D}$ and $i > 0;\\
0,          & $otherwise;$\\
\end{array}
\right .
\end{equation*}
where $\overline{D} = 2^s - 1$, and $s$ is the size of a $D_i$
sequence element in bits.\\

The $J_k$ sequence will be defined recursively in the following
way:
\begin{equation*}
J_k = \left \lbrace
\begin{array}{ll}
L_0, & $if $k = 0;\\
L_j, & $otherwise where $j = \min \lbrace i \ | \ \Delta L_i >
\overline{D}$
and $L_i > J_{k-1} \rbrace.\\
\end{array}
\right .
\end{equation*}
Here the $D_i$ sequence is called the overflow difference
sequence. There is an obvious distinction between $\Delta L_i$ and
$D_i$, but the latter will also be called the difference sequence,
if it is not too disturbing. $J_k$ it is called the jump sequence.
The compression method which makes use of the $D_i$ and $J_k$
sequences will be called difference sequence compression (DSC).
The $D_i$ and $J_k$ sequences together will be called the DSC
header. \hfill $\square$
\end{Def}

Notice here that $\Delta L_i$ and $D_i$ are basically the same
sequence. The only difference is that some elements of the
original difference sequence $\Delta L_i$ are replaced with zeros,
if and only if they cannot be stored in $s$ bits. (The symbol $s$
denotes a natural number. The theoretically optimal value of $s$
can be determined, if the distribution of $\Delta L_i$ is known.
In practice, for performance reasons, $s$ is either 8 or 16 or
32.)

The difference sequence will also be called the relative logical
position sequence, and we shall call the jump sequence the
absolute logical position sequence.

From the definitions of $D_i$ and $J_k$, one can see clearly that,
for every zero element of the $D_i$ sequence, there is exactly one
corresponding element in the $J_k$ sequence. For example, let us
assume that $D_0 = D_3 = D_5 = 0$, and $D_1, D_2, D_4, D_6, D_7,
D_8 > 0$. Then the above mentioned correspondence is shown in the
following table:

\begin{table}[h]
\begin{center}
\begin{tabular}{@{\extracolsep{\fill}}|c|c|c|c|c|c|c|c|c|c|}
    \hline
    $D_0$ & $D_1$ & $D_2$ & $D_3$ & $D_4$ & $D_5$ & $D_6$ & $D_7$ & $D_8$ & \dots \cr
    \hline
    $J_0$ &       &       & $J_1$ &       & $J_2$ &       &       &       & \dots \cr
    \hline
\end{tabular}
\end{center}
\end{table}

From the above definition, the recursive formula below follows for
$L_j$.
\begin{equation*}
L_j = \left \lbrace
\begin{array}{ll}
L_{j-1} + D_j, & $if $D_j > 0;\\
J_k,           & $otherwise where $k = \min \lbrace i \ | \ J_i >
L_{j-1}
\rbrace.\\
\end{array}
\right .
\end{equation*}
In other words, every element of the $L_j$ sequence can be
calculated by adding zero or more consecutive elements of the
$D_i$ sequence to the proper
jump sequence element. For instance, in the above example\\
\\
$L_0 = J_0$;\\
$L_1 = J_0 + D_1$;\\
$L_2 = J_0 + D_1 + D_2$;\\
$L_3 = J_1$;\\
$L_4 = J_1 + D_4$;\\
and so on.\\

A detailed analysis of DSC and the search algorithm can be found
in \cite{SzepkutiD}.

\emph{Difference\,--\,Huffman coding.} The key idea in
difference\,--\,Huffman coding is that we can compress the
difference sequence further if we replace it with its
corresponding Huffman code.

\begin{Def} The compression method, which uses the jump
sequence ($J_k$) and the Huffman code of the difference sequence
($D_i$), will be labelled difference\,--\,Huffman coding (DHC).
The $J_k$ sequence and the Huffman code of the $D_i$ sequence
together will be called the DHC header. \hfill $\square$
\end{Def}

The difference sequence usually contains a lot of \emph{zeros}.
Moreover, it contains many \emph{ones} too if there are numerous
consecutive elements in the $L_j$ sequence of logical positions.
By definition, the elements of the difference sequence are smaller
than those of the logical position sequence. The elements of $D_j$
will recur with greater or less frequency. Hence it seems
reasonable to code the frequent elements with fewer bits, and the
less frequent ones with more. To do this, the optimal prefix code
can be determined by the well-known Huffman algorithm
\cite{Huffman}.

\section{A Model Based on the Dominance of the I/O Cost Rule}
\label{sec:IOCost}

During our analysis of caching effects, we followed two different
approaches:

\begin{itemize}
    \item For the first model, we applied the \emph{dominance of the I/O cost} rule
          to calculate the expected number of I/O operations.
    \item In the second one, instead of counting the number of disk
          inputs/outputs, we introduced two different constants: $D_m$
          and $D_t$. The constant $D_m$ denotes the time needed to
          retrieve one cell from the disk, if the multidimensional
          representation is used. The constant $D_t$ shows the time required
          to read one row from the disk, if the table representation is
          used. The constants were determined experimentally.
          The tests showed that $D_m \ll D_t$, that is more disk I/O operations are needed to
          retrieve one row from the table representation than one cell
          from the multidimensional representation which is obvious when there is no
          caching. However, for the second model, it was not necessary
          to compute the exact number of I/O operations for the
          alternative physical representations due to the
          experimental approach.
\end{itemize}

The first model is described in this section, whereas the second
model in the next one.

Throughout the paper, we suppose that the different database pages
are accessed with the same probability. In other words, uniform
distribution will be assumed.

It is not hard to see that this assumption corresponds to the
worst case. If the distribution is not uniform, then certain
partitions of the pages will be read/written with higher
probability than the average. Therefore it is more likely to find
pages from these partitions in the buffer cache than from other
parts of the database. Hence the non-uniform distribution
increases the buffer hit rate and thus the performance.

We are going to estimate the number of database pages (blocks) in
the buffer cache. First it will be done for the multidimensional
representation, then for the table representation.

\emph{Multidimensional physical representation.} In this paper, we
shall assume that prefetching is not performed by the system.
Hence, for the multidimensional representation, one or zero
database page has to be copied from the disk into the memory, when
a cell is accessed. This value is one if the needed page is not in
the buffer cache, zero otherwise.

The multidimensional representation requires that the header and
the dimension values are preloaded into the memory. The total size
of these will be denoted by $H$. The compressed multidimensional
array can be found on the disk. The pages of the latter are
gradually copied into the memory as a result of caching. Thus the
total memory occupancy of this representation can be computed by
adding $H$ to the size of the buffer cache.

\begin{Def} In this section, for the multidimensional
representation, we shall use the following notation.\\
$N$ is the number of pages required to store the compressed array ($N \geqq 1$);\\
$B_i$ is the expected value of the number of pages in the buffer
cache after the $i^\mathrm{th}$ database access ($i \geqq 0$).
\hfill $\square$
\end{Def}

\begin{Thm} \label{thm:Multidimensional} Suppose that $B_k$ is less than the size of the
memory\footnote{Please note that the memory size is also measured
in pages in this section.} available for caching for every $k \in
\{0, 1, \dots, i\}$ index. In addition, let us assume that the
buffer cache is `cold' initially, \emph{i.e.} $B_0 = 0$. Then, for
the multidimensional representation,
\begin{equation*}
  B_i = N \left( 1 - \left( 1 - \frac{1}{N} \right) ^ i \right).
\end{equation*}
\begin{proof} The theorem will be proven by induction.
For convenience, let us define $d$ as follows:
\begin{equation*}
  d = 1 - \frac{1}{N}.
\end{equation*}
For $i = 0$, the theorem holds:\footnote{We define $0^0$ as 1. In
this way, the theorem remains true for the special case of $N =
1$.}
\begin{equation*}
  B_0 = N \left( 1 - \left( 1 - \frac{1}{N} \right) ^ 0 \right)
      = N \left( 1 - d ^ 0 \right)
      = N \left( 1 - 1 \right) = 0.
\end{equation*}
Now assume that the theorem has already been proven for $i - 1$:
\begin{equation*}
  B_{i - 1} = N \left( 1 - d ^ {i - 1} \right).
\end{equation*}
Then for $i$ we obtain that
\begin{equation*}
  B_i = B_{i - 1} + 0 \times \frac{B_{i - 1}}{N}
      + 1 \times \frac{N - B_{i - 1}}{N}.
\end{equation*}
Because of the uniform distribution, $\frac{B_{i - 1}}{N}$ is the
probability that the required database block can be found in the
memory. Zero new page will be copied from the disk into the buffer
cache in this situation. However, in the opposite case, one new
page will be brought into the memory. This will occur with
probability $\frac{N - B_{i - 1}}{N}$. In other words, the
expected value of the increase is
\begin{equation}\label{eq:Increase}
  0 \times \frac{B_{i - 1}}{N} + 1 \times \frac{N - B_{i - 1}}{N}
  = \frac{N - B_{i - 1}}{N}
  = 1 - \frac{B_{i - 1}}{N}.
\end{equation}
Hence
\begin{equation*}
  B_i = B_{i - 1} + 1 - \frac{B_{i - 1}}{N}
      = B_{i - 1} \left( 1 - \frac{1}{N} \right) + 1
      = B_{i - 1} d + 1.
\end{equation*}
From the induction hypothesis follows that
\begin{equation*}
  B_i = N \left( 1 - d ^ {i - 1} \right) d + 1.
\end{equation*}
It is easy to see that
\begin{equation*}
  B_i = 1 + d + d ^ 2 + d ^ 3 + \dots + d ^ {i - 1}
      = \frac{1 - d ^ i}{1 - d}
      = N \left( 1 - d ^ i \right).
\end{equation*}
The last formula can be written as
\begin{equation*}
  B_i = N \left( 1 - \left( 1 - \frac{1}{N} \right) ^ i \right),
\end{equation*}
which proves the theorem.
\end{proof}
\end{Thm}

The time to retrieve one cell from the multidimensional
representation is proportional to the number of pages brought into
the memory. The latter is a linear function of the size of the
buffer cache. This is rephrased in the following theorem.

\begin{Thm} \label{thm:LinearMultidimensional} Assume that the number of database pages in the buffer
cache is $B$. The memory available for caching is greater than
$B$. Let us suppose that a cell is accessed in the
multidimensional representation. Then the expected number of pages
copied from the disk into the memory is
\begin{equation*}
  1 - \frac{B}{N}.
\end{equation*}
\begin{proof} Similarly to Equation (\ref{eq:Increase}), the expected
number of pages necessary for this operation is
\begin{equation*}
  0 \times \frac{B}{N} + 1 \times \frac{N - B}{N}
  = \frac{N - B}{N}
  = 1 - \frac{B}{N}.
\end{equation*}
\end{proof}
\end{Thm}

\begin{Rem} The above theorem holds even if $B$ is equal to the
number of pages available for caching. However, in this case, the
database management system (or the operating system) has to remove
a page from the buffer cache, if a page fault happens. If the
removed page is `dirty,' then it has to be written back to the
disk in order not to lose the modifications. That is why another
disk I/O operation is needed. In this paper, we are going to
ignore these situations, because most OLAP applications handle the
data in a \emph{read only} or \emph{read mostly} way.
\end{Rem}

\begin{figure}
\caption{\label{fig:array}The expected number of pages copied from
the disk into the memory, if the multidimensional representation
is used}
\begin{center}
\resizebox{100mm}{!}{\rotatebox{-90}{\includegraphics{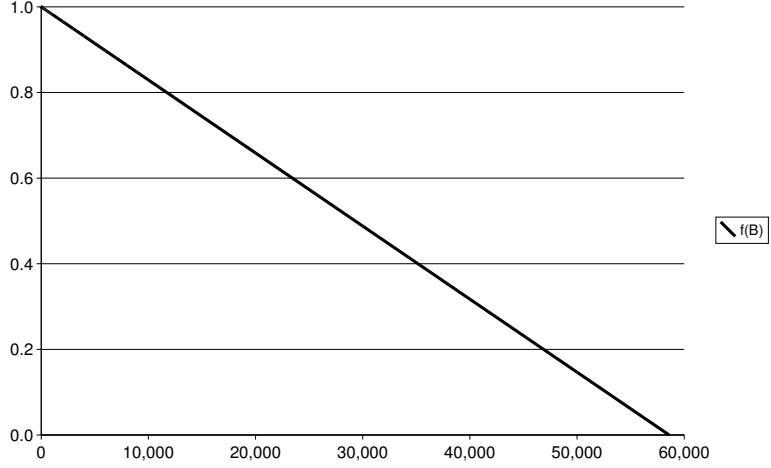}}}
\end{center}
\end{figure}

Figure \ref{fig:array} illustrates the behaviour of the
multidimensional representation. The horizontal axis shows the
number of pages in the buffer cache. The vertical one demonstrates
the expected number of pages retrieved from the disk. The $f(B)$
function is defined as follows:
\begin{equation*}
  f(B) = 1 - \frac{B}{N}.
\end{equation*}

\emph{Table physical representation.} Now, let us turn to the
other storage method, the table representation. Both the table and
B-tree index are kept on the disk. The table itself could be
handled similarly to the compressed array, but the B-tree index is
structured differently. It consists of several levels. In our
model, we are going to consider these levels separately. To
simplify the notation, the table will also be considered as a
separate level. The following definition introduces the necessary
notations.

\begin{Def} \label{def:Bi}
$L \geqq 2$ is the number of levels in the table representation.
On level 1, the root page of the B-tree can be found. Level $L -
1$ is the last level of the B-tree, which contains the leaf nodes.
Level $L$ corresponds to the table.\\
$N_\ell \geqq 1$ is the number of pages on level $\ell$ ($1 \leqq
\ell \leqq L$). Specifically, $N_1 = 1$, as there is only one root
page.

The total number of pages is
\begin{equation} \label{eq:N}
N = \sum_{\ell = 1}^{L}{N_\ell}.
\end{equation}
$B_i^{(\ell)} \geqq 0$ is the number of pages in the buffer cache
from level $\ell$ after the $i^\mathrm{th}$ database access ($1
\leqq \ell \leqq L$ and $i \geqq 0$).\\
The total number of pages in the buffer cache is
\begin{equation} \label{eq:Bi}
B_i = \sum_{\ell = 1}^{L}{B_i^{(\ell)}}.
\end{equation}
\hfill $\square$
\end{Def}

\begin{Thm} \label{thm:Bi} Suppose that $B_k$ is less than the size of the memory
available for caching for every $k \in \{0, 1, \dots, i\}$ index.
In addition, let us assume that the buffer cache is cold
initially: $B_0 = 0$. Then, for the table representation,
\begin{equation*}
  B_i = N - \sum_{\ell = 1}^L{N_\ell \left( 1 - \frac{1}{N_\ell} \right) ^ i}.
\end{equation*}
\begin{proof} Observe that we can apply the result of Theorem
\ref{thm:Multidimensional} at each level:
\begin{equation} \label{eq:Bil}
  B_i^{(\ell)} = N_\ell \left( 1 - \left( 1 - \frac{1}{N_\ell} \right) ^ i \right)
               = N_\ell  - N_\ell \left( 1 - \frac{1}{N_\ell} \right) ^
               i.
\end{equation}
The assertion of the theorem follows from the definitions of $N$
and $B_i$ shown in Equations (\ref{eq:N}) and (\ref{eq:Bi}):
\begin{equation*}
  B_i = \sum_{\ell = 1}^L{B_i^{(\ell)}}
      = \sum_{\ell = 1}^L{\left( N_\ell - N_\ell \left( 1 - \frac{1}{N_\ell} \right) ^ i \right)},
\end{equation*}
\begin{equation*}
  B_i = \sum_{\ell = 1}^L{N_\ell} - \sum_{\ell = 1}^L{N_\ell \left( 1 - \frac{1}{N_\ell} \right) ^ i},
\end{equation*}
\begin{equation*}
  B_i = N - \sum_{\ell = 1}^L{N_\ell \left( 1 - \frac{1}{N_\ell} \right) ^ i}.
\end{equation*}
\end{proof}
\end{Thm}

Similarly to the other representation, the necessary time to
retrieve one row from the table representation is proportional to
the number of pages brought into the memory. The next theorem
investigates how the number of pages brought into the memory
depends on the size of the buffer cache.

\begin{Thm} Assume that the number of database pages in the buffer
cache is $B_i = \sum_{\ell = 1}^L{B_i^{(\ell)}}$. The memory
available for caching is greater than $B_i$. Let us suppose that a
row is accessed in the table representation. Then the expected
number of pages read from the disk into the memory is
\begin{equation} \label{eq:LinearWithLVars}
  L - \sum_{\ell = 1}^L{\frac{B_i^{(\ell)}}{N_\ell}}.
\end{equation}
\begin{proof} This will be shown by applying the result of Theorem
\ref{thm:LinearMultidimensional} per level. For level $\ell$, the
number of pages copied into the memory is:
\begin{equation*}
  1 - \frac{B_i^{(\ell)}}{N_\ell}.
\end{equation*}
Hence, for all levels in total, it is:
\begin{equation*}
  \sum_{\ell = 1}^L{\left( 1 - \frac{B_i^{(\ell)}}{N_\ell} \right)}
  = \sum_{\ell = 1}^L{1} - \sum_{\ell = 1}^L{\frac{B_i^{(\ell)}}{N_\ell}}
  = L - \sum_{\ell = 1}^L{\frac{B_i^{(\ell)}}{N_\ell}}.
\end{equation*}
\end{proof}
\end{Thm}

$L, N_1, N_2, \dots, N_L$ are constants. Therefore Equation
(\ref{eq:LinearWithLVars}) is a linear function of $B_i^{(1)},
B_i^{(2)}, \dots, B_i^{(L)}$. The same expression can be looked at
as a function of $B_i$, as well:

\begin{Def} \label{def:f}
\begin{equation*}
  f(B_i) = L - \sum_{\ell = 1}^L{\frac{B_i^{(\ell)}}{N_\ell}}.
\end{equation*} \hfill $\square$
\end{Def}

Just like before, we are going to assume that the buffer cache is
cold initially: $B_0 = 0$. If this is the case, then $B_0^{(\ell)}
= 0$ for every $\ell \in \{ 1, 2, \dots, L \}$, because of
Definition \ref{def:Bi}. Therefore,
\begin{equation*}
  f(B_0) = L - \sum_{\ell = 1}^L{\frac{0}{N_\ell}} = L.
\end{equation*}
In other words, one page per level has to be read into the memory
at the first database access. If the memory available for caching
is not smaller than $L$, then $B_1^{(\ell)} = 1$ for every $\ell$
and
\begin{equation*}
  B_1 = \sum_{\ell = 1}^L{B_1^{(\ell)}}
      = \sum_{\ell = 1}^L{1} = L.
\end{equation*}
Obviously, we obtain the same, if we use the alternative
(recursive) formula:
\begin{equation*}
  B_1 = B_0 + f(B_0) = 0 + L = L.
\end{equation*}

Now, let us investigate the special case, when $N_m = \max \{ N_1,
N_2, \dots, N_L \} = 1.$ Because of the latter, there is only one
page per level ($N_1 = N_2 = \dots = N_\ell = 1$), which means
that $N$ also equals $L$. To put it into another way, the entire
database is cached into the memory after the first database
access, given that the available memory is greater than or equal
to the size of the database. After this, there is no need to copy
more pages into the memory:
\begin{equation*}
  f(B_1) = L - \sum_{\ell = 1}^L{\frac{B_1^{(\ell)}}{N_\ell}}
         = L - \sum_{\ell = 1}^L{\frac{1}{1}} = L - L = 0.
\end{equation*}
To summarise this paragraph, below we show the values of $B_i$ and
$f(B_i)$ for every $i$:
\begin{eqnarray*}
     B_0 &=& 0,                                   \\
     B_1 &=& B_2 = \dots = B_i = \dots = L,       \\
  f(B_0) &=& L,                                   \\
  f(B_1) &=& f(B_2) = \dots = f(B_i) = \dots = 0. \\
\end{eqnarray*}
In the remainder of this section, we shall assume that $N_m > 1$.

For sufficiently large $i$ values, $f(B_i)$ can be considered a
linear function of $B_i$. This is the main idea behind the theorem
below.

\begin{Thm} \label{thm:LinearTable} Suppose that $B_k$ is less than the size of the memory
available for caching for every $k \in \{0, 1, \dots, i\}$ index.
In addition, let us assume that $B_0 = 0$, $B_i < N$ and $f(B_i)
\neq 0$. Then, for the table representation,
\begin{equation*}
  f(B_i) \rightarrow \frac{N - B_i}{N_m}, \textrm{ if } i \rightarrow \infty,
\end{equation*}
where $N_m = \max \{ N_1, N_2, \dots, N_L \}$.
\begin{proof}
First, we show that
\begin{equation*}
  f(B_i) = \frac{N - B_i}{W_i},
\end{equation*}
where $W_i$ is a weighted average of constants $N_1, N_2, \dots,
N_L$. Then we demonstrate that $W_i$ tends to $N_m$, if $i$ tends
to infinity. From Equation (\ref{eq:Bil}), we know that
\begin{equation*}
  \frac{B_i^{(\ell)}}{N_\ell} =
  \frac{N_\ell - N_\ell \left( 1 - \frac{1}{N_\ell} \right) ^ i}
       {N_\ell} =
  1 - \left( 1 - \frac{1}{N_\ell} \right) ^ i.
\end{equation*}
Using Definition \ref{def:f}, we obtain that
\begin{equation*}
  f(B_i)
  = L - \sum_{\ell = 1}^L{\frac{B_i^{(\ell)}}{N_\ell}}
  = L - \sum_{\ell = 1}^L{\left( 1 - \left( 1 - \frac{1}{N_\ell} \right) ^ i
  \right)}
  = \sum_{\ell = 1}^L{\left( 1 - \frac{1}{N_\ell} \right) ^ i}.
\end{equation*}
Theorem \ref{thm:Bi} implies the following equation:
\begin{equation*}
  N - B_i
  = \sum_{\ell = 1}^L{N_\ell \left( 1 - \frac{1}{N_\ell} \right) ^
  i}.
\end{equation*}
Let us define $W_i$ as follows:
\begin{equation*}
  W_i = \frac
  {\sum_{\ell = 1}^L{N_\ell \left( 1 - \frac{1}{N_\ell} \right) ^i}}
  {\sum_{\ell = 1}^L{\left( 1 - \frac{1}{N_\ell} \right) ^ i}},
\end{equation*}
given that the denominator is not zero ($f(B_i) \neq 0$). Observe
that $W_i$ is a weighted average of constants $N_1, N_2, \dots,
N_L$. The weight of $N_\ell$ is $\left( 1 - \frac{1}{N_\ell}
\right) ^i$ for every $\ell \in \{1, 2, \dots, L\}$. With the
previous definition, we get that
\begin{equation*}
  W_i = \frac{N - B_i}{f(B_i)}.
\end{equation*}
If $W_i$ does not vanish ($B_i < N$), then
\begin{equation*}
  f(B_i) = \frac{N - B_i}{W_i}.
\end{equation*}
Finally, we have to prove that $W_i \rightarrow N_m$, if $i
\rightarrow \infty$. For every $\ell \in \{1, 2, \dots L\}$, the
inequality $1 \leqq N_\ell \leqq N_m$ holds. It is not difficult
to see that
\begin{equation} \label{eq:NlLessThanNm}
  \frac{\left( 1 - \frac{1}{N_\ell} \right)^i}{\left( 1 - \frac{1}{N_m} \right)^i}
  \rightarrow 0, \textrm{ if } N_\ell < N_m \textrm{ and } i \rightarrow \infty.
\end{equation}
Obviously
\begin{equation} \label{eq:NlEqualsNm}
  \frac{\left( 1 - \frac{1}{N_\ell} \right)^i}{\left( 1 - \frac{1}{N_m} \right)^i}
  = 1, \textrm{ if } N_\ell = N_m > 1.
\end{equation}
From Equations (\ref{eq:NlLessThanNm}) and (\ref{eq:NlEqualsNm}),
it follows immediately, that
\begin{equation*}
  W_i = \frac
  {\sum_{\ell = 1}^L{N_\ell \left( 1 - \frac{1}{N_\ell} \right) ^i}}
  {\sum_{\ell = 1}^L{\left( 1 - \frac{1}{N_\ell} \right) ^ i}}
  = \frac
  {\sum_{\ell = 1}^L{N_\ell
    \frac{\left( 1 - \frac{1}{N_\ell} \right) ^ i}
         {\left( 1 - \frac{1}{N_m} \right) ^ i}
  }}
  {\sum_{\ell = 1}^L{
    \frac{\left( 1 - \frac{1}{N_\ell} \right) ^ i}
         {\left( 1 - \frac{1}{N_m} \right) ^ i}
  }}
  \rightarrow N_m, \textrm{ if } i \rightarrow \infty.
\end{equation*}
\end{proof}
\end{Thm}

\begin{figure}
\caption{\label{fig:table}The expected number of pages copied from
the disk into the memory, if the table representation is used}
\begin{center}
\resizebox{100mm}{!}{\rotatebox{-90}{\includegraphics{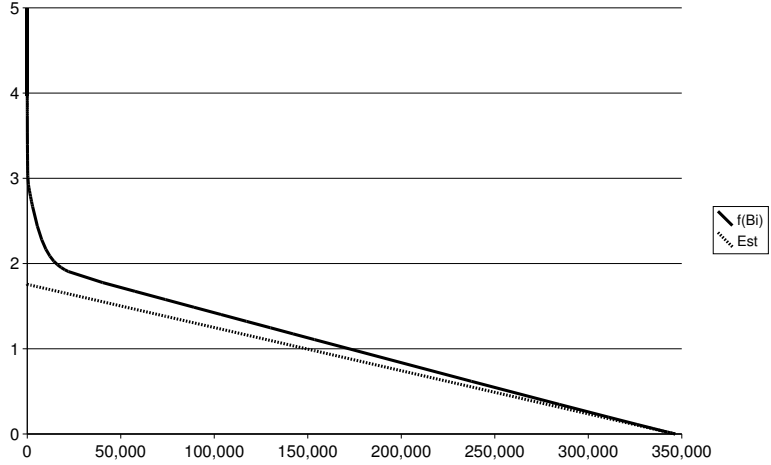}}}
\end{center}
\end{figure}

Figure \ref{fig:table} demonstrates the behaviour of the table
representation. The horizontal axis is the number of pages in the
buffer cache. The vertical one shows the expected number of pages
retrieved from the disk. The Estimation denoted by `Est.' in the
chart is the limit of the $f(B_i)$ function:
\begin{equation*}
  \textit{Estimation} = \frac{N - B_i}{N_m}.
\end{equation*}

We conclude this section by summarising the findings:
\begin{itemize}
  \item If we assume requests with uniform distribution, then the
        expected number of database pages brought into the memory
        at a database access is a linear function of the number of
        pages in the buffer cache.
  \item Specifically, for the multidimensional representation, it
        equals
        \begin{equation*}
          1 - \frac{B}{N},
        \end{equation*}
        where $B$ is the number of pages in the buffer cache and
        $N$ is the size of the compressed multidimensional array
        in pages.
  \item For the table representation, it is
        \begin{equation*}
          f(B_i) = L - \sum_{\ell = 1}^L{\frac{B_i^{(\ell)}}{N_\ell}},
        \end{equation*}
        where $L$ is the number of levels, $B_i^{(\ell)}$ is the
        number of pages in the buffer cache from level $\ell$,
        $B_i = \sum_{\ell = 1}^L{B_i^{(\ell)}}$ and
        $N_\ell$ is the total number of pages on level $\ell$.
  \item The expression above is a linear function of $B_i^{(1)},
        B_i^{(2)}, \dots, B_i^{(L)}$, but for large $i$ values, it
        can be considered as a linear function of $B_i$, as well,
        because
        \begin{equation*}
          f(B_i) \rightarrow \frac{N - B_i}{N_m}, \textrm{ if } i \rightarrow \infty,
        \end{equation*}
        where $N_m = \max \{ N_1, N_2, \dots, N_L \}$ and
        $N = \sum_{\ell = 1}^L{N_\ell}$.
\end{itemize}

\section{An Alternative Model}
\label{sec:Alternative}

In this section we shall examine how the caching affects the speed
of retrieval in the different physical database representations.
For the analysis, a model will be proposed. Then we will give
sufficient and necessary conditions for such cases where the
expected retrieval time is smaller in one representation than in
the other.

The caching can speed up the operation of a database management
system significantly if the same block is requested while it is
still in the memory. In order to show how the caching modifies the
results of this paper, let us introduce the following notations.

\begin{Def}
\begin{eqnarray*}
  M   & = & \textrm{the retrieval time, if the information is in the memory,} \\
  D   & = & \textrm{the retrieval time, if the disk also has to be accessed,} \\
  p   & = & \textrm{the probability of having everything needed in the memory,} \\
  q   & = & 1 - p, \\
  \xi & = & \textrm{how long it takes to retrieve the requested information.}
\end{eqnarray*} \hfill $\square$
\end{Def}

In our model we shall consider $M$ and $D$ constants. Obviously,
$\xi$ is a random variable. Its expected value can be calculated
as follows:
\begin{equation*}
\mathbb{E}(\xi) = pM + qD.
\end{equation*}

Notice that $D$ does not tell us how many blocks have to be read
from the disk. This also means that the value of $D$ will be
different for the table and the multidimensional representations.
The reason for this is that, in general, at most one block has to
be read with the multidimensional representation. Exactly one
reading is necessary if nothing is cached, because only the
compressed multidimensional array is kept on the disk. Everything
else (the header, the dimension values, and so forth) is loaded
into the memory in advance. With the table representation, more
block readings may be needed because we also have to traverse
through the B-tree first, and then we have to retrieve the
necessary row from the table.

$M$ is also different for the two alternative physical
representations. This is because two different algorithms are used
to retrieve the same information from two different physical
representations.

Hence, for the above argument, we are going to introduce four
constants.

\begin{Def}
\begin{eqnarray*}
  \hspace{1.25em} M_m & = & \textrm{the value of $M$ for the multidimensional representation,} \\
                  M_t & = & \textrm{the value of $M$ for the table representation,} \\
                  D_m & = & \textrm{the value of $D$ for the multidimensional representation,} \\
                  D_t & = & \textrm{the value of $D$ for the table representation.} \\
                      &   & \hspace{27.75em} \square
\end{eqnarray*}
\end{Def}

If we sample the cells/rows with uniform probability\footnote{In
this section, just like in the previous one, we shall make the
same assumption that every cell/row is sampled with the same
probability.}, we can then estimate the probabilities as follows:
\begin{eqnarray*}
  p & = & \frac{\textrm{the number of cached pages}}
               {\textrm{the total size in pages}}, \\
  q & = & 1 - p.
\end{eqnarray*}
By the `total size' we mean that part of the physical
representation which can be found on the disk at the beginning. In
the multidimensional representation, it is the compressed
multidimensional array, whereas in the table representation, we
can put the entire size of the physical representation into the
denominator of $p$. The cached pages are those that had been
originally on the disk, but were moved into the memory later. In
other words, the size of the cached blocks (numerator) is always
smaller than or equal to the total size (denominator).

The experiments show that the alternative physical representations
differ from each other in size. That is why it seems reasonable to
introduce four different probabilities in the following manner.

\begin{Def}
\begin{eqnarray*}
  p_m & = & \textrm{the value of $p$ for the multidimensional representation,} \\
  p_t & = & \textrm{the value of $p$ for the table representation,} \\
  q_m & = & 1 - p_m, \\
  q_t & = & 1 - p_t.
\end{eqnarray*} \hfill $\square$
\end{Def}

When does the inequality below hold? This is an important
question:
\begin{equation*}
\mathbb{E}(\xi_m) < \mathbb{E}(\xi_t).
\end{equation*}
Here $\xi_m$ and $\xi_t$ are random variables that are the
retrieval times in the multidimensional and table representations,
respectively.

In our model, $\mathbb{E}(\xi_i) = p_i M_i + q_i D_i$ ($i \in
\lbrace m, t \rbrace$). Thus the question can be rephrased as
follows:
\begin{equation*}
p_m M_m + q_m D_m < p_t M_t + q_t D_t.
\end{equation*}

The value of the $M_m$, $D_m$, $M_t$ and $D_t$ constants was
measured by carrying out some experiments. (See the following
section.) Two different results were obtained. For one benchmark
database (TPC-D), the following was found:
\begin{equation*}
M_t < M_m \ll D_m \ll D_t.
\end{equation*}
Another database (APB-1) gave a slightly different result:
\begin{equation*}
M_m \ll M_t \ll D_m \ll D_t.
\end{equation*}

The $M_m \ll D_m$ and $M_t \ll D_m$ inequalities hold because disk
operations are slower than memory operations by orders of
magnitude. The third one ($D_m \ll D_t$) is because we have to
retrieve more blocks from the table representation than from the
multidimensional to obtain the same information.

Note here that $\mathbb{E}(\xi_i)$ is the convex linear
combination of $M_i$ and $D_i$ ($p_i, q_i \in [0, 1]$ and $i \in
\lbrace m, t \rbrace$). In other words, $\mathbb{E}(\xi_i)$ can
take any value from the closed interval $[M_i, D_i]$.

The following provides sufficient condition for $\mathbb{E}(\xi_m)
< \mathbb{E}(\xi_t)$:
\begin{equation*}
D_m < p_t M_t + q_t D_t.
\end{equation*}
From this, we can obtain the inequality constraint:
\begin{eqnarray*}
D_m & < & p_t M_t + (1 - p_t) D_t, \\
p_t & < & \frac{D_t - D_m}{D_t - M_t}.
\end{eqnarray*}

The value for $\frac{D_t - D_m}{D_t - M_t}$ was found to be
63.2\%, 66.5\% and 66.3\% (for TPC-D, TPC-H and APB-1,
respectively) in the experiments. This means that, based on the
experimental results, the expected value of the retrieval time was
smaller in the multidimensional representation than in the table
representation when less than 63.2\% of the latter one was cached.
This was true regardless of the fact whether the multidimensional
representation was cached or not.

Now we are going to distinguish two cases based on the value of
$M_m$ and $M_t$.

\emph{Case 1: $M_t < M_m$.} This was true for the TPC-D benchmark
database. (Here the difference sequence consisted of 16-bit
unsigned integers, which resulted in a slightly more complicated
decoding, as the applied Huffman decoder returns 8 bits at a time.
This may be the reason why $M_m$ became larger than $M_t$.) In
this case, we can give a sufficient condition for
$\mathbb{E}(\xi_m) > \mathbb{E}(\xi_t)$, as the equivalent
transformations below show:
\begin{eqnarray*}
p_t M_t + q_t D_t           & < & M_m, \\
p_t M_t + (1 - p_t) D_t     & < & M_m, \\
\frac{D_t - M_m}{D_t - M_t} & < & p_t.
\end{eqnarray*}

For $\frac{D_t - M_m}{D_t - M_t}$ we obtained a value of 99.9\%.
This means that the expected retrieval time was smaller in the
\emph{table} representation when more than 99.9\% of it was
cached. This was true even when the whole multidimensional
representation was in the memory.

\emph{Case 2: $M_m < M_t$.} This inequality holds for the TPC-H
and the APB-1 benchmark databases. Here we can give another
sufficient condition for $\mathbb{E}(\xi_m) < \mathbb{E}(\xi_t)$:
\begin{eqnarray*}
p_m M_m + q_m D_m           & < & M_t, \\
p_m M_m + (1 - p_m) D_m     & < & M_t, \\
\frac{D_m - M_t}{D_m - M_m} & < & p_m.
\end{eqnarray*}

The left hand side of the last inequality was equal to 99.9\% and
98.3\% for the TPC-H and APB-1 benchmark databases, respectively.
In other words, when more than 99.9\% of the multidimensional
representation was cached, it then resulted in a faster operation
on average than the table representation regardless of the caching
level of the latter.

Finally, let us give a necessary and sufficient condition for
$\mathbb{E}(\xi_m) < \mathbb{E}(\xi_t)$. First, let us consider
the following equivalent transformations (making the natural
assumption that $D_t > M_t$):
\begin{eqnarray}
  \mathbb{E}(\xi_m)       & < & \mathbb{E}(\xi_t),                 \label{eq:EV}   \\
  p_m M_m + q_m D_m       & < & p_t M_t + q_t D_t,                                 \\
  p_m M_m + (1 - p_m) D_m & < & p_t M_t + (1 - p_t) D_t,                           \\
  p_t                     & < & \frac{D_m - M_m}{D_t - M_t} p_m +
                                \frac{D_t - D_m}{D_t - M_t}.       \label{eq:pt}
\end{eqnarray}
The last inequality was the following for the three tested
databases, TPC-D, TPC-H and APB-1, respectively:
\begin{eqnarray*}
  p_t & < & 0.368 p_m + 0.632, \\
  p_t & < & 0.335 p_m + 0.665, \\
  p_t & < & 0.343 p_m + 0.663.
\end{eqnarray*}

\begin{Thm} Suppose that $D_t > M_t$. Then the expected retrieval time is
smaller in the case of the multidimensional physical
representation than in the table physical representation if and
only if
\begin{equation*}
  p_t < \frac{D_m - M_m}{D_t - M_t} p_m + \frac{D_t - D_m}{D_t - M_t}.
\end{equation*}

\begin{proof}
The truth of the theorem is a direct consequence of Equations
(\ref{eq:EV})\,--\,(\ref{eq:pt}).
\end{proof}
\end{Thm}

Now, let us change our model slightly. In this modified version,
we shall assume that the different probabilities are (piecewise)
linear functions of the memory size available. This assumption is
in accordance with Theorems \ref{thm:LinearMultidimensional} and
\ref{thm:LinearTable}. With the multidimensional representation,
the formula below follows from the model for the expected
retrieval time:
\begin{equation*}
T_m(x) = M_m p_m(x) + D_m q_m(x) = M_m p_m(x) + D_m (1 - p_m(x)),
\end{equation*}
\begin{equation*}
T_m(x) = (M_m - D_m) p_m(x) + D_m,
\end{equation*}
where
\begin{equation*}
p_m(x) = \min \left \lbrace \frac{x - H}{C}, 1 \right \rbrace,
\end{equation*}
$H$ is the total size of the multidimensional representation part,
which is loaded into the memory in advance (the header and the
dimension values), $C$ is the size of the compressed
multidimensional array and $x$ ($\geqq H$) is the size of the
available memory.

In an analogous way, for the table representation, we obtain the
formula:
\begin{equation*}
T_t(x) = M_t p_t(x) + D_t q_t(x) = M_t p_t(x) + D_t (1 - p_t(x)),
\end{equation*}
\begin{equation*}
T_t(x) = (M_t - D_t) p_t(x) + D_t,
\end{equation*}
where
\begin{equation*}
p_t(x) = \min \left \lbrace \frac{x}{S}, 1 \right \rbrace,
\end{equation*}
$S$ is the total size of the table representation and $x$ ($\geqq
0$) is the size of the memory available for caching.

It is not hard to see that the global maximum and minimum values
and locations of the functions $T_m(x)$ and $T_t(x)$ are the
following:
\begin{displaymath}
\begin{array}{ccccc}
  \max \{ T_m(x) \ | \ x \geqq H \} = D_m & $and$ & T_m(x) = D_m & $if and only if$ & x = H,         \cr
                                                                                                     \cr
  \min \{ T_m(x) \ | \ x \geqq H \} = M_m & $and$ & T_m(x) = M_m & $if and only if$ & x \geqq H + C, \cr
                                                                                                     \cr
  \max \{ T_t(x) \ | \ x \geqq 0 \} = D_t & $and$ & T_t(x) = D_t & $if and only if$ & x = 0,         \cr
                                                                                                     \cr
  \min \{ T_t(x) \ | \ x \geqq 0 \} = M_t & $and$ & T_t(x) = M_t & $if and only if$ & x \geqq S.
\end{array}
\end{displaymath}

\begin{Def} For $x \geqq H$ values, let us define the speed-up factor in the following way:
\begin{equation*}
  \textit{speed-up}(x) = \frac{T_t(x)}{T_m(x)}.
\end{equation*} \hfill $\square$
\end{Def}

The global maximum of the speed-up factor can be achieved, when
the entire multidimensional representation is cached into the
memory. This is specified in the following theorem.

\begin{Thm} Suppose that
\begin{equation} \label{eq:a1b1Anda2b2}
  0 > \frac{M_t - D_t}{S} > \frac{M_m - D_m}{C}
  \textrm{\ \ \ and\ \ \ }
  0 < -\frac{M_m - D_m}{C} H + D_m < D_t.
\end{equation}
Then the global maximum of the $\textit{speed-up}(x)$ function can
be found at $C + H$.
\begin{proof} The $\textit{speed-up}(x)$ function is continuous, because
$T_t(x)$ and $T_m(x)$ are continuous and $T_m(x) \neq 0$. Hence,
to prove the theorem, it is enough to show that this function is
strictly monotone increasing on interval $(H,\ C + H)$, strictly
monotone decreasing on $(C + H,\ S)$ and constant on $(S,\
\infty)$. On the first interval,
\begin{equation*}
  \textit{speed-up}(x)
  = \frac{(M_t - D_t) p_t(x) + D_t}
         {(M_m - D_m) p_m(x) + D_m}
  = \frac{(M_t - D_t) \frac{x}{S} + D_t}
         {(M_m - D_m) \frac{x - H}{C} + D_m}.
\end{equation*}
For convenience, let us introduce the following notation:
\begin{eqnarray*}
  a_1 &=& \frac{M_t - D_t}{S},          \\
  b_1 &=& D_t,                          \\
  a_2 &=& \frac{M_m - D_m}{C},          \\
  b_2 &=& -\frac{M_m - D_m}{C} H + D_m.
\end{eqnarray*}
The first derivative of the $\textit{speed-up}(x)$ function is
\begin{equation*}
  \textit{speed-up}'(x)
  = \left( \frac{a_1 x + b_1}{a_2 x + b_2} \right)'
  = \frac{a_1 b_2 - a_2 b_1}{(a_2x + b_2)^2}.
\end{equation*}
The first derivative is positive if and only if $a_1 b_2 - a_2 b_1
> 0$. Equation (\ref{eq:a1b1Anda2b2}) can be written as
\begin{equation} \label{eq:a1GreaterThana2}
  0 > a_1 > a_2
\end{equation}
and
\begin{equation} \label{eq:b2LessThanb1}
  0 < b_2 < b_1.
\end{equation}
Let us multiply Equation (\ref{eq:a1GreaterThana2}) by $b_1$,
Equation (\ref{eq:b2LessThanb1}) by $a_1$. Then we obtain that
\begin{equation*}
  a_1 b_1 > a_2 b_1
\end{equation*}
and
\begin{equation*}
  a_1 b_2 > a_1 b_1.
\end{equation*}
From the last two inequalities, we get that $a_1 b_2 > a_2 b_1$,
which is equivalent with $a_1 b_2 - a_2 b_1 > 0$. Thus
$\textit{speed-up}'(x) > 0$ and $\textit{speed-up}(x)$ is strictly
monotone increasing on interval $(H,\ C + H)$.

Now, suppose that $x \in (C + H,\ S)$. In this case
\begin{equation*}
  \textit{speed-up}(x)
  = \frac{(M_t - D_t) p_t(x) + D_t}
         {M_m}
  = \frac{(M_t - D_t) \frac{x}{S} + D_t}
         {M_m}
  = \frac{a_1 x + b_1}{M_m}.
\end{equation*}
The fist derivative is
\begin{equation*}
  \textit{speed-up}'(x) = \frac{a_1}{M_m} < 0,
\end{equation*}
because $a_1 < 0$ and $M_m > 0$. So $\textit{speed-up}(x)$ is
strictly monotone decreasing.

Finally, let us take the case, when $x \in (S,\ \infty)$. The
speed-up factor
\begin{equation*}
  \textit{speed-up}(x) = \frac{M_t}{M_m},
\end{equation*}
which is constant.
\end{proof}
\end{Thm}

The location of the global maximum is $C + H$. The global maximum
value is obviously
\begin{equation*}
  \textit{speed-up}(C + H)
  = \frac{a_1 (C + H) + b_1}{M_m}
  = \frac{\frac{M_t - D_t}{S} (C + H) + D_t}{M_m}.
\end{equation*}
As it will be described in details in the next section,
experiments were made to determine the value of the constants. For
these data, see Table \ref{tab:Constants} there. The sizes were
also measured and can be seen in Table \ref{tab:Speed-up} (in
bytes) together with the global maximum locations and values per
benchmark database. As it can be seen from the latter table, the
speed-up can be very large, 2\,--\,3 orders of magnitude. The
maximum value for the TPC-D benchmark database was more than 400,
while for the APB-1 benchmark database, it was more than 1,500.

\begin{table}[h]
\caption{\label{tab:Speed-up}Global maximum of
$\textit{speed-up}(x)$}
\begin{center}
\begin{tabular}{@{\extracolsep{\fill}}l|rrr}
  \hline
  \bf Symbol                 & \bf TPC-D   & \bf TPC-H     & \bf APB-1     \cr
  \hline
  $S$                        & 279,636,324 & 1,419,181,908 & 1,295,228,960 \cr
  $C$                        &  48,007,720 &   239,996,040 &    99,144,000 \cr
  $H$                        &  19,006,592 &   154,024,844 &     4,225,039 \cr
  $C + H$                    &  67,014,312 &   394,020,884 &   103,369,039 \cr
  $\textit{speed-up}(C + H)$ &         416 &         1,066 &         1,549 \cr
  \hline
\end{tabular}
\end{center}
\end{table}

We can draw the conclusions of this section as follows:
\begin{itemize}
  \item If (nearly) the entire physical representation is cached
        into the memory, then the complexity of the algorithm will
        determine the speed of retrieval. A less CPU-intensive
        algorithm will result in a faster operation.
  \item In the tested cases, the expected retrieval time was
        smaller with multidimensional physical representation
        when less than 63.2\% of the table representation was
        cached. This was true regardless of the caching
        level of the multidimensional representation.
  \item Depending on the size of the memory available for caching,
        the speed-up factor can be very large, up to 2\,--\,3 orders of
        magnitude! In other words, the caching effects of the
        alternative physical representations modify the results
        significantly. Hence these effects should always be
        taken into account, when the retrieval time of the
        different physical representations are compared with each
        other.
\end{itemize}

\section{Experiments}
\label{sec:Experiments}

We carried out experiments in order to measure the sizes of the
different physical representations and the constants in the
previous section. We also examined how the size of the cache
influenced the speed of retrieval.

Table \ref{tab:HW and SW} shows the hardware and software used for
testing. The speed of the processor, the memory and the hard disk
all influence the experimental results quite significantly, just
like the memory size. In the computer industry, all of these
parameters have increased quickly over the time. But the increase
of the hard disk speed has been somewhat slower. Hence, it is
expected that the results presented will remain valid despite the
continuing improvement in computer technology.

\begin{table}[h]
\caption{\label{tab:HW and SW}Hardware and software used for
testing}
\begin{center}
\begin{tabular}{@{\extracolsep{\fill}}ll}
\hline
  Processor            & Intel Pentium 4 with HT technology, 2.6 GHz,   \cr
                       & 800 MHz FSB, 512 KB cache                      \cr
  Memory               & 512 MB, DDR 400 MHz                            \cr
  Hard disk            & Seagate Barracuda, 80 GB, 7200 RPM, 2 MB cache \cr
  Filesystem           & ReiserFS format 3.6 with standard journal      \cr
  Page size of B-tree  & 4 KB                                           \cr
  Operating system     & SuSE Linux 9.0 (i586)                          \cr
  Kernel version       & 2.4.21-99-smp4G                                \cr
  Compiler             & gcc (GCC) 3.3.1 (SuSE Linux)                   \cr
  Programming language & C                                              \cr
  Free                 & procps version 3.1.11                          \cr
\hline
\end{tabular}
\end{center}
\end{table}

In the experiments we made use of three benchmark databases: TPC-D
\cite{TPC-D}, TPC-H \cite{TPC-H} and APB-1 \cite{APB-1}. One
relation ($R$) was derived per benchmark database in exactly the
same way as was described in \cite{SzepkutiC}. Then these
relations were represented physically with a multidimensional
representation and table representation.

Tables \ref{tab:TPC-D}, \ref{tab:TPC-H} and \ref{tab:APB-1} show
that DHC results in a smaller multidimensional representation than
difference sequence compression. (For TPC-H, the so-called Scale
Factor was equal to 5. That is why the table representation of
TPC-H is about five times greater than that of TPC-D.)

\begin{table}[h]
\caption{\label{tab:TPC-D}TPC-D benchmark database}
\begin{center}
\begin{tabular*}{\hsize}{@{\extracolsep{\fill}}l|rr}
  \hline
  \bf Compression                 & \bf Size in bytes & \bf Percentage\cr
  \hline
  \bf Table representation                                            \cr
  Uncompressed                    &      279,636,324 &         100.0\%\cr
  \hline
  \bf Multidimensional representation                                 \cr
  Difference sequence compression &       67,925,100 &          24.3\%\cr
  Difference\,--\,Huffman coding  &       67,014,312 &          24.0\%\cr
  \hline
\end{tabular*}
\end{center}
\end{table}

\begin{table}[h]
\caption{\label{tab:TPC-H}TPC-H benchmark database}
\begin{center}
\begin{tabular*}{\hsize}{@{\extracolsep{\fill}}l|rr}
  \hline
  \bf Compression                 & \bf Size in bytes & \bf Percentage\cr
  \hline
  \bf Table representation                                            \cr
  Uncompressed                    &     1,419,181,908 &        100.0\%\cr
  \hline
  \bf Multidimensional representation                                 \cr
  Difference sequence compression &       407,414,614 &          28.7\%\cr
  Difference\,--\,Huffman coding  &       394,020,884 &          27.8\%\cr
  \hline
\end{tabular*}
\end{center}
\end{table}

\begin{table}[h]
\caption{\label{tab:APB-1}APB-1 benchmark database}
\begin{center}
\begin{tabular*}{\hsize}{@{\extracolsep{\fill}}l|rr}
  \hline
  \bf Compression                 & \bf Size in bytes & \bf Percentage\cr
  \hline
  \bf Table representation                                            \cr
  Uncompressed                    &     1,295,228,960 &        100.0\%\cr
  \hline
  \bf Multidimensional representation                                 \cr
  Difference sequence compression &       113,867,897 &          8.8\%\cr
  Difference\,--\,Huffman coding  &       103,369,039 &          8.0\%\cr
  \hline
\end{tabular*}
\end{center}
\end{table}

In the rest of this section, we shall deal only with DHC. Its
performance will be compared to the performance of the
uncompressed table representation.

In order to determine the constant values of the previous section,
an experiment was performed. A random sample was taken with
replacement from relation $R$ with uniform distribution. The
sample size was 1000. Afterwards the sample elements were
retrieved from the multidimensional representation and then from
the table representation. The elapsed time was measured to
calculate the average retrieval time per sample element. Then the
same sample elements were retrieved again from the two physical
representations. Before the first round, nothing was cached. So
the results help us to determine the constants $D_m$ and $D_t$.
Before the second round, every element of the sample was cached in
both physical representations. So the times measured in the second
round correspond to the values of the constants $M_m$ and $M_t$.
The results of the experiment can be seen in Table
\ref{tab:Constants}.

\begin{table}[h]
\caption{\label{tab:Constants}Constants}
\begin{center}
\begin{tabular}{@{\extracolsep{\fill}}l|rrr}
  \hline
  \bf        & \bf TPC-D & \bf TPC-H & \bf APB-1 \cr
  \bf Symbol & \bf  (ms) & \bf  (ms) & \bf  (ms) \cr
  \hline
  $M_m$      &     0.031 &     0.014 &     0.012 \cr
  $M_t$      &     0.021 &     0.018 &     0.128 \cr
  $D_m$      &     6.169 &     7.093 &     6.778 \cr
  $D_t$      &    16.724 &    21.165 &    19.841 \cr
  \hline
\end{tabular}
\end{center}
\end{table}

\begin{figure}
\caption{\label{fig:TPC-D}The retrieval time for the TPC-D
benchmark database as a function of the memory size available for
caching}
\begin{center}
\resizebox{100mm}{!}{\includegraphics{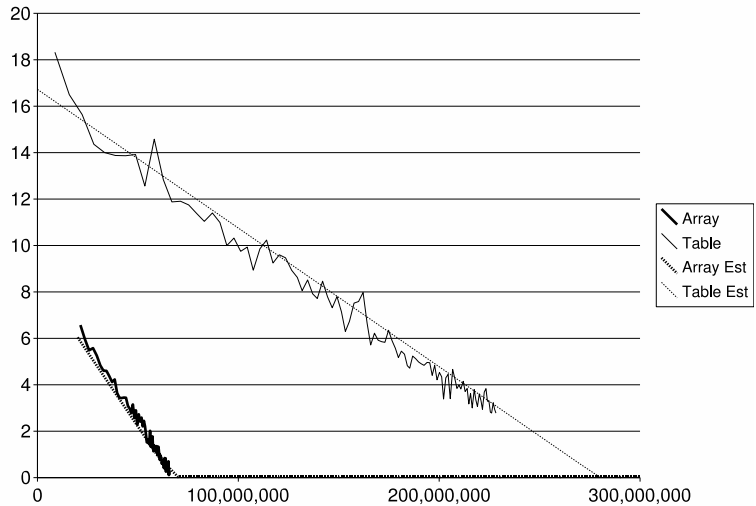}}
\end{center}
\end{figure}

\begin{figure}
\caption{\label{fig:TPC-H}The retrieval time for the TPC-H
benchmark database as a function of the memory size available for
caching}
\begin{center}
\resizebox{100mm}{!}{\rotatebox{-90}{\includegraphics{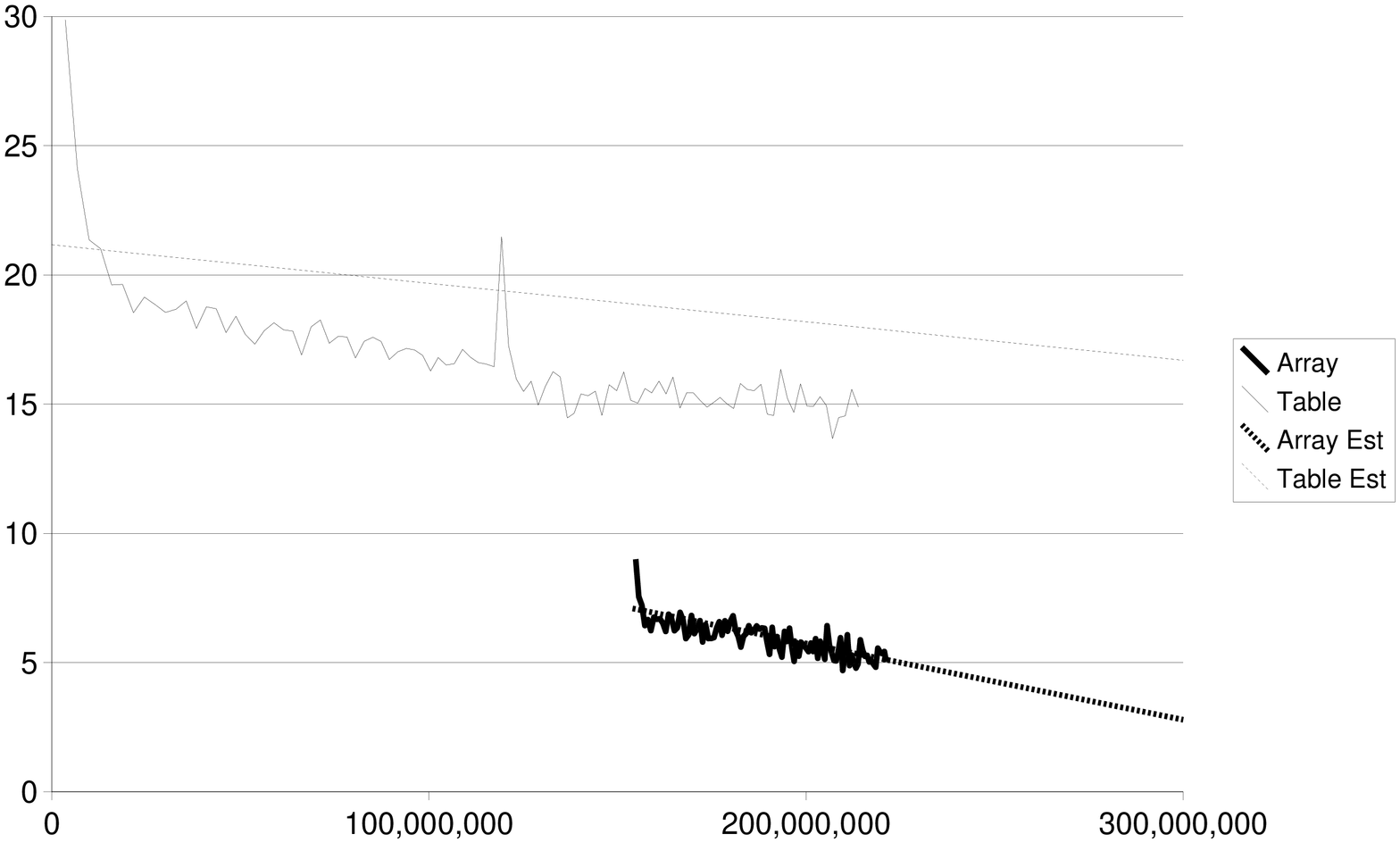}}}
\end{center}
\end{figure}

\begin{figure}
\caption{\label{fig:APB-1}The retrieval time for the APB-1
benchmark database as a function of the memory size available for
caching}
\begin{center}
\resizebox{100mm}{!}{\includegraphics{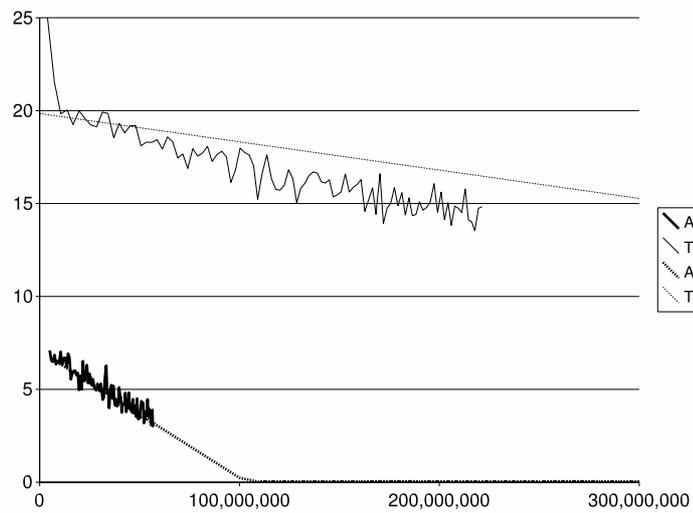}}
\end{center}
\end{figure}

In the next experiment, we examined how the size of memory
available for caching influenced the speed of retrieval. In
Figures \ref{fig:TPC-D}, \ref{fig:TPC-H} and \ref{fig:APB-1},
$T_m(x)$ is labelled as `Array Est.,' $T_t(x)$ as `Table Est.' The
horizontal axis shows the size of the memory in bytes, while the
vertical one displays the expected/average retrieval time in
milliseconds.

In order to verify the model with empirical data, we performed the
following tests. Random samples were taken with replacement. The
sample size was set at 300 in TPC-D and 100 in TPC-H and APB-1 in
order to stay within the constraints of the physical memory. The
average retrieval time was measured as well as the cache size used
for each physical representation. In the multidimensional
representation, the utilized cache size was corrected by adding
$H$ to it, as this representation requires that some parts of it
are loaded into the memory in advance. Then the above sampling and
measuring procedures were repeated another 99 times. That is,
altogether 30,000 elements were retrieved from the TPC-D database,
and 10,000 from TPC-H and APB-1. The average retrieval time, as a
function of the cache size (or memory) used, can also be seen in
Figures \ref{fig:TPC-D}\,--\,\ref{fig:APB-1}. The data relating to
the multidimensional physical representation are labelled as
`Array,' and the data for the table representation as `Table.'

The diagrams suggest that the model fits the empirical data quite
well. Only the table representation of TPC-H and ABP-1 deviates
slightly from it.


The test results of the first ten passes and the last ten passes
can be seen in Tables \ref{tab:Retrieval1} and
\ref{tab:Retrieval2}, as well. Column A is the sequence number.
Columns B\,--\,E correspond to TPC-D, columns F\,--\,I to TPC-H,
while columns J\,--\,M are for APB-1. Columns B, F and J show the
memory needed for the multidimensional representation, while
columns C, G and K give the same for the table representation. The
retrieval time with the multidimensional representation can be
found in columns D, H and L, and the table representation in
columns E, I and M. The `memory used' values are strictly
increasing. This can be attributed to the fact that increasingly
larger parts of the physical representations are cached into the
memory.

Looking at Tables \ref{tab:Retrieval1}\,--\,\ref{tab:Retrieval2}
and Figures \ref{fig:TPC-D}\,--\,\ref{fig:APB-1}, it can be seen
that the multidimensional representation was always significantly
faster over the tested range.

\begin{table}[h]
\caption{\label{tab:Retrieval1}Memory used (in $2^{10}$ bytes) and
retrieval time (in milliseconds) for the TPC-D and TPC-H benchmark
databases}
\begin{center}
\begin{tabular}{@{\extracolsep{\fill}}r|rrrr|rrrr}
  \hline
       A &      B &       C &      D &      E &      F  &       G &      H &      I \cr
  \hline
       1 & 20,893 &   8,500 &   6.57 &  18.32 & 151,215 &   3,524 &   9.00 &  29.86 \cr
       2 & 23,093 &  15,488 &   5.96 &  16.50 & 152,015 &   6,644 &   7.54 &  24.10 \cr
       3 & 25,097 &  21,732 &   5.48 &  15.64 & 152,811 &   9,684 &   7.21 &  21.36 \cr
       4 & 27,025 &  27,420 &   5.58 &  14.36 & 153,591 &  12,652 &   6.43 &  21.01 \cr
       5 & 28,841 &  32,668 &   5.26 &  14.00 & 154,367 &  15,528 &   6.66 &  19.61 \cr
       6 & 30,565 &  37,896 &   4.83 &  13.88 & 155,139 &  18,328 &   6.23 &  19.63 \cr
       7 & 32,113 &  42,908 &   4.61 &  13.87 & 155,919 &  21,160 &   6.75 &  18.54 \cr
       8 & 33,557 &  47,684 &   4.60 &  13.92 & 156,707 &  23,992 &   6.67 &  19.14 \cr
       9 & 34,949 &  52,228 &   4.37 &  12.56 & 157,463 &  26,760 &   6.70 &  18.85 \cr
      10 & 36,289 &  56,792 &   4.12 &  14.58 & 158,231 &  29,456 &   6.53 &  18.55 \cr
  \vdots & \vdots &  \vdots & \vdots & \vdots &  \vdots &  \vdots & \vdots & \vdots \cr
      91 & 63,609 & 216,352 &   0.35 &   2.94 & 211,143 & 193,868 &   5.28 &  15.78 \cr
      92 & 63,677 & 217,228 &   0.70 &   3.69 & 211,683 & 195,556 &   5.02 &  14.93 \cr
      93 & 63,729 & 218,060 &   0.24 &   3.83 & 212,235 & 197,240 &   5.07 &  14.91 \cr
      94 & 63,769 & 218,784 &   0.22 &   3.29 & 212,795 & 198,940 &   4.93 &  15.29 \cr
      95 & 63,813 & 219,484 &   0.28 &   3.31 & 213,359 & 200,584 &   4.82 &  14.95 \cr
      96 & 63,841 & 220,200 &   0.34 &   2.82 & 213,895 & 202,164 &   5.56 &  13.67 \cr
      97 & 63,857 & 220,804 &   0.13 &   2.78 & 214,439 & 203,760 &   5.42 &  14.48 \cr
      98 & 63,905 & 221,592 &   0.30 &   3.23 & 215,019 & 205,464 &   5.34 &  14.54 \cr
      99 & 63,925 & 222,260 &   0.11 &   2.94 & 215,583 & 207,140 &   5.43 &  15.57 \cr
     100 & 63,949 & 222,908 &   0.32 &   2.78 & 216,099 & 208,864 &   5.03 &  14.89 \cr
  \hline
\end{tabular}
\end{center}
\end{table}

\begin{table}[h]
\caption{\label{tab:Retrieval2}Memory used (in $2^{10}$ bytes) and
retrieval time (in milliseconds) for the APB-1 benchmark database}
\begin{center}
\begin{tabular}{@{\extracolsep{\fill}}r|rrrr}
  \hline
       A &      J &       K &      L &      M \cr
  \hline
       1 &  4,926 &   3,840 &   7.10 &  24.99 \cr
       2 &  5,698 &   7,204 &   6.55 &  21.53 \cr
       3 &  6,478 &  10,312 &   6.48 &  19.83 \cr
       4 &  7,262 &  13,452 &   6.85 &  20.03 \cr
       5 &  8,002 &  16,328 &   6.35 &  19.25 \cr
       6 &  8,774 &  19,336 &   6.52 &  19.99 \cr
       7 &  9,506 &  22,208 &   6.42 &  19.56 \cr
       8 & 10,266 &  25,076 &   7.02 &  19.23 \cr
       9 & 10,978 &  27,884 &   6.35 &  19.13 \cr
      10 & 11,726 &  30,664 &   6.68 &  19.92 \cr
  \vdots & \vdots &  \vdots & \vdots & \vdots \cr
      91 & 52,334 & 201,140 &   3.72 &  13.82 \cr
      92 & 52,726 & 202,836 &   4.46 &  14.86 \cr
      93 & 53,046 & 204,540 &   3.55 &  14.75 \cr
      94 & 53,438 & 206,240 &   3.98 &  14.52 \cr
      95 & 53,754 & 207,960 &   3.47 &  15.77 \cr
      96 & 54,090 & 209,516 &   3.82 &  14.12 \cr
      97 & 54,382 & 211,100 &   3.09 &  14.01 \cr
      98 & 54,670 & 212,660 &   3.13 &  13.53 \cr
      99 & 55,054 & 214,404 &   3.89 &  14.74 \cr
     100 & 55,358 & 216,144 &   2.97 &  14.83 \cr
  \hline
\end{tabular}
\end{center}
\end{table}

Summarizing our experimental results, we may say that:
\begin{itemize}
    \item The size of DHC was smaller than that of the difference
          sequence compression.
    \item With suitably designed experiments, we were able to
          measure the constants of the model proposed in the
          previous section.
    \item We verified the model with empirical data.
    \item Over the tested range of available memory, the
          multidimensional representation was always much quicker
          than the table representation in terms of retrieval time.
\end{itemize}

\section{Conclusion}
\label{sec:Conclusion}

It often turns out that caching significantly improves response
times. This was also found to be the case for us when the same
relation was represented physically in different ways. In order to
analyse this phenomenon, we proposed two models.

In the first model, the \emph{dominance of the I/O cost rule} was
used to examine the caching effects. Uniform distribution was
assumed for the analysis. We found that the expected number of
pages brought into the memory is a linear function of the buffer
cache size. And we know that the time to retrieve a cell/row from
the database is proportional to the number of database pages
copied from the disk into the memory.

The second model was built in accordance with the findings of the
first one. In the latter model, four constants were introduced for
the retrieval time from the memory ($M_m$ and $M_t$) and from the
disk ($D_m$ and $D_t$). It was necessary to have four symbols as
we had to distinguish between the multidimensional representation
($M_m$ and $D_m$) and the table representation ($M_t$ and $D_t$).
Based on the model, necessary and sufficient conditions were given
for when one physical representation results in a lower expected
retrieval time than the other. Actually, with the tested benchmark
databases, we found that the expected retrieval time was smaller
with a multidimensional physical representation if less than
63.2\% of the table representation was cached. This was true
regardless of the caching level of the multidimensional
representation.

We were able to infer from the second model that the complexity of
the algorithm could determine the speed of retrieval when (nearly)
the entire physical representation was cached into the memory. A
less CPU-intensive algorithm will probably result in a faster
operation. It is important to mention that the first model is
unable to explain this phenomenon. The reason for this is that the
\emph{dominance of the I/O cost rule} ignores the time
requirements of the memory operations.

Using a slightly modified version of the second model, we
investigated the speed-up factor, which can be achieved, if the
multidimensional representation is used instead of the table one.
We found that, depending on the memory size available for caching,
the speed-up can be 2\,--\,3 orders of magnitude. That is why it
is very important to also take into account the caching effects,
when the performances of the different physical representations
are compared.

Experiments were performed to measure the constants of the model.
We found that there was a big difference in values between $M_m$
and $M_t$, as well as $D_m$ and $D_t$. The difference of the first
two constants can be accounted for by the different CPU-intensity
of the algorithms. The reason why $D_m \ll D_t$ is that the
multidimensional representation requires much less I/O operations
than the table representation when one cell/row is retrieved. This
latter observation is in line with the \emph{dominance of the I/O
cost} rule. However, instead of counting the number of I/O
operations, we chose to determine the values of $D_m$ and $D_t$
from empirical data.

We verified the model with additional experiments and found that
the model fitted the experimental results quite well. There was
only a slight difference with the table representation of the
TPC-H and APB-1 benchmark databases.

Finally, over the tested range of available memory, the
multidimensional representation was always much faster than the
table representation in terms of average retrieval time, as it can
be seen in Figures \ref{fig:TPC-D}\,--\,\ref{fig:APB-1}.

Based on the above results, we think, like {\sc Westmann} \emph{et
al.}\ \cite{Westmann}, that today's database systems should be
extended with compression capabilities to improve their overall
performance.

\section*{Acknowledgments}

I would like to thank Prof.\ Dr.\ J\'anos Csirik for his
continuous support and very useful suggestions.

\end{document}